\documentclass[aps,superscriptaddress,twocolumn,prb]{revtex4}
\usepackage{amsmath}
\usepackage{bm}
\usepackage{epsfig}

\newcommand{\cL}{{\cal L}}
\newcommand{\cA}{{\cal A}}
\newcommand{\cB}{{\cal B}}
\newcommand{\cG}{{\cal G}}
\newcommand{\cU}{{\cal U}}

\newcommand{\ti}{\tilde}

\newcommand{\nl}{\nonumber \\}
\newcommand{\la}{\langle}
\newcommand{\ra}{\rangle}
\newcommand{\Sec}[1]{Sec.\,\ref{#1}}

\newcommand{\be}{\begin{equation}}
\newcommand{\ee}{\end{equation}}
\newcommand{\bea}{\begin{eqnarray}}
\newcommand{\eea}{\end{eqnarray}}
\newcommand{\bsube}{\begin{subequations}}
\newcommand{\esube}{\end{subequations}}

\newcommand{\Eq}[1]{Eq.\,(\ref{#1})}
\newcommand{\Eqs}[1]{Eqs.\,(\ref{#1})}
\newcommand{\Fig}[1]{Fig.\,\ref{#1}}
\newcommand{\Figs}[1]{Figs.\,\ref{#1}}

\newcommand{\B}{\mbox{\tiny B}}

\newcommand{\HTA}{\mbox{\tiny HTA}}

\begin{document}

\title{Hierarchical quantum master equation with
 semiclassical Drude dissipation}
\author{Rui-Xue Xu}  \email{rxxu@ustc.edu.cn}
\author{Bao-Ling Tian}
\author{Jian Xu}
\affiliation{Hefei National Laboratory for Physical Sciences at
Microscale, University of Science and Technology of China, Hefei,
Anhui 230026, China}

\affiliation{Department of Chemistry, Hong Kong University of
Science and Technology, Kowloon, Hong Kong SAR, China}

\author{Qiang Shi} \email{qshi@iccas.ac.cn}
\affiliation{Beijing National Laboratory for Molecular Sciences,
State Key Laboratory for Structural Chemistry of Unstable and Stable Species,
Institute of Chemistry, Chinese Academy
of Sciences, Beijing 100190, China}

\author{YiJing Yan} \email{yyan@ust.hk}
\affiliation{Department of Chemistry, Hong Kong University of
Science and Technology, Kowloon, Hong Kong SAR, China}

\date{\today}

\begin{abstract}
   We propose a nonperturbative
quantum dissipation theory, in term of
hierarchical quantum master equation.
It may be used with
a great degree of confidence to various dynamics
systems in condensed phases.
The theoretical development is rooted
in an improved semiclassical
treatment of Drude bath, beyond the
conventional high temperature approximations.
It leads to the new theory a simple modification  but important
improvement over the conventional stochastic
Liouville equation theory, without extra numerical cost.
Its broad range of validity and applicability
is extensively
demonstrated with two--level electron transfer model systems,
where the new theory can be considered as the modified
Zusman equation.
We also present a criterion, which
depends only on the system--bath coupling strength,
characteristic bath memory time, and temperature,
to estimate the performance of the hierarchical quantum master equation.

\end{abstract}
\maketitle

\section{Introduction}
\label{thintro}

Dissipation is often inevitable and plays important
roles in many systems. The key quantity
in quantum dissipation theory (QDT)
is the reduced system density operator,
$\rho(t)\equiv {\rm tr}_{\B} \rho_{\rm tot}(t)$;
i.e., the trace of total composite one over bath subspace.
For Gaussian bath, exact QDT can be
formulated with path integral\cite{Fey63118,Wei08,Kle09} or its differential
version in terms of
hierarchical equations of motion (HEOM).\cite{
Tan06082001,Tan89101,Tan906676,Tan914131,Tan9466,Ish053131,%
Xu05041103,Xu07031107,Jin08234703,Zhe09164708,Shi09084105,Shi09164518,Yan04216,Zho08034106}
However, exact approaches are numerically expensive in general.
In this paper, we propose an approximate HEOM theory,
which will be termed hereafter
as hierarchical quantum master equation (HQME).
Compared with the traditional high--temperature approximation (HTA)
schemes such as the stochastic Liouville
equation,\cite{Tan06082001,Kub63174,Kub69101,Abr092350,Zhu093750}
the new method
requires about the same numerical effort,
but enjoys a greatly improved range of
validity.
Moreover, the HQME supports
also a convenient and versatile criterion of applicability
that seems to be rather insensitive
to specific  systems.

 We will exemplify the validity
and applicability of the present theory
with the standard electron transfer
spin--boson model system.
In this case, the HQME is analytically solvable,
and its high--temperature limit is just
the celebrated Zusman equation
(ZE).\cite{Zus80295,Zus8329,Gar854491,Yan89281}
The ZE treats the effect of bath via a diffusive solvation coordinate.
Its validity has been a subject of study for
years.\cite{Fra97427,Fra992075,Tho012991,Jun023822,%
Muh03179,Ank041436,Dod06257,Zha0311864,Zha049630}
With the aid of analytical solutions,
we can now readily exploit the range of validity
for both HQME and ZE,
over the full parameters space
of electron transfer systems.

The present development is rooted
in an improved semiclassical treatment of
the fluctuation--dissipation theorem.
Consider a Gaussian stochastic bath variable $F_{\B}(t)$.
It can be the fluctuating solvation coordinate  $U(t)-\la U\ra_{\B}$
in electron transfer systems,
or the fluctuating transition frequency
$\delta\omega_{eg}(t)-\la\delta\omega_{eg}\ra_{\B}$
in spectroscopy.
Its effect on system is completely described by
the bath correlation
function, $C(t) \equiv \la F_{\B}(t)F_{\B}(0)\ra_{\B}$.
The classical Gaussian--Markovian description assumes
\be\label{Ctcl}
   C_{\rm cl}(t) = 2\lambda k_BT e^{-\gamma t}.
\ee
Adopted is the classical fluctuation--dissipation relation,
such as
$\la U^2\ra_{\B}-\la U\ra_{\B}^2\approx 2k_BT\la U\ra_{\B}
=2\lambda k_BT$.
This model has been widely used, for example, in the spectroscopic
motional narrowing problems.\cite{Kub85,Kub63174,Kub69101}
It also leads to the stochastic Liouville
equation description of reduced system
dynamics.\cite{Tan06082001,Kub63174,Kub69101,Abr092350,Zhu093750}
However, it discards the imaginary part that is responsible
for the spectroscopic Stokes shift or solvent reorganization.
In other words, the classical bath correlation function does not
consider the back action of system on bath.
 The conventional HTA scheme adopts
\be\label{CtHTA}
   C_{\HTA}(t) =   \lambda (2k_BT - i\gamma) e^{-\gamma t}.
\ee
Here, the imaginary part assumes no approximation
in the Drude dissipation model, while
the real part remains its classical form.
This scheme does account for
the back action of system on bath.
However, the reduced system density
matrix dynamics based on $C_{\HTA}(t)$
encounters rather often the positivity
violation problem
that was originally not suffered by that
based on classical $C_{\rm cl}(t)$.
%

The above two conventional schemes are
the low--order approximations
of fluctuation--dissipation theorem, together
with the Drude bath spectral density model,
\be \label{DrudeJ}
    J(\omega) = \frac{2\lambda\gamma\omega}{\omega^2 + \gamma^2} \, .
\ee
The exact fluctuation--dissipation theorem reads\cite{Wei08}
\be \label{FDT}
   C(t) 
   =  \frac{1}{\pi}\int_{-\infty}^{\infty}\!\!d\omega\, e^{-i\omega t}
      \frac{J(\omega)}{1-e^{-\beta\omega}}\,,
\ee
with $\beta=1/(k_BT)$ being the inverse temperature.
Consider the bosonic function in the expansion:
\be\label{FDTexpan}
   \frac{1}{1-e^{-\beta\omega}}=\frac{1}{\beta\omega}+\frac{1}{2}
   +\frac{\beta\omega}{12}+{\cal O}[(\beta\omega)^3].
\ee
The classical $C_{\rm cl}(t)$ [\Eq{Ctcl}] uses
only the lowest order expansion;
the high--temperature $C_{\HTA}(t)$ [\Eq{CtHTA}] includes also
the second term.

 The proposed HQME approach is based
on the following semiclassical bath correlation
function,
\be\label{CtMZE}
  C_{\rm sc}(t) = C_{\HTA}(t)
   - \frac{\lambda\gamma}{6k_BT}[\gamma e^{-\gamma t}-2\delta(t)].
\ee
It uses all the three lowest terms of \Eq{FDTexpan},
thus, improving over $C_{\HTA}(t)$ by two orders
in $\beta\omega$.
Interestingly, the resulting HQME for reduced dynamics
costs almost no additional numerical effort,
but largely overcomes
the aforementioned positivity problem.
This remarkable feature
will be exemplified in simple electron transfer systems,
where the HTA limit of HQME  has been
shown to be equivalent to the ZE.\cite{Shi09164518}

The remainder of paper is organized as follows.
We present the HQME on the basis of $C_{\rm sc}(t)$ [\Eq{CtMZE}],
its stochastic Liouville
equation description,
and its continued fraction Green's function theory,
respectively, in the three subsections of \Sec{theory}.
The exact HEOM formalism based on $C(t)$ of \Eq{FDT} is briefed in Appendix.
In \Sec{ana}, we consider the two--level electron transfer
spin--boson model, in which the continued fraction
Green's function theory of HQME can be
analytically resolved and also the ZE is recovered
in the HTA limit.
Exemplified with this model,
numerical studies on validity and applicability
of HQME are carried out in \Sec{num}.
We show that the HQME remarkably
improves over its HTA/ZE counterpart, in terms of
both positivity and electron transfer dynamics.
Comments and discussions about
the criterion
for the applicability of HQME to arbitrary systems
are presented in \Sec{thcomm}.
Finally, we conclude the paper in \Sec{sum}.

\section{Theory}
\label{theory}
\subsection{Hierarchical quantum master equation}
\label{scheom}
 Consider the total Hamiltonian in the form of
\be\label{HTstart}
  H_{\rm T}(t) = H+Q F_{\B}(t).
\ee
Denote ${\cal L}\hat O=[H,\hat O]$
as the reduced system Liouvillian,
and set $\hbar=1$ throughout this paper.
The last term of \Eq{HTstart} is the system--bath
coupling, in which the system operator $Q$ defines the dissipative mode,
through which the stochastic bath operator
or generalized Langevin force $F_{\B}(t)$
acting on the system.
Recast the semiclassical bath correlation function of \Eq{CtMZE} as
\be\label{Ct0}
  C_{\rm sc}(t)=(c_r-ic_i)e^{-\gamma t}+2\Delta\delta(t),
\ee
where
\be\label{cDel}
  c_r =2\lambda k_BT-\gamma\Delta,
\ \ \
  c_i=\lambda\gamma,
\ \ \
 \Delta =\frac{\lambda\gamma}{6k_BT}\,.
\ee
The corresponding HQME can then be
constructed via the standard
calculus--on--path--integral algebra.\cite{Tan89101,Xu05041103,Xu07031107}
It reads
\begin{align} \label{modZEheom}
\dot\rho_n (t)
 &=-(i{\cal L}+\delta{\cal R}+n\gamma) \rho_n(t)
\nl&\quad
   -i\sqrt{n}{\cal A}\rho_{n-1}(t)-i\sqrt{n+1}\,{\cal B}\rho_{n+1}(t) ,
\end{align}
with
\bsube
\begin{align}
  &\delta{\cal R}\hat O =\Delta[Q,[Q,\hat O]]\,,
 \ \ \ \
 {\cal B}\hat O =  \sqrt{|c_r|}\,[Q, \hat O]\,,
\\
 &{\cal A}\hat O = \Bigl( c_r[Q,\hat O] - ic_i\{Q,\hat O\} \Bigr)/\sqrt{|c_r|}\,.
\end{align}
\esube
In this formalism, the reduced system
density operator, $\rho(t)\equiv\rho_0(t)$, couples
hierarchically with a set of
well--defined auxiliary density operators
$\{\rho_{n>0}(t)\}$. The hierarchical construction
resolves not just
system--bath coupling strengths but also memory time scales.
Each $\rho_{n}$, having been scaled by the factor $(n!|c_r|^n)^{-1/2}$,
if compared with Ref.\,\onlinecite{Xu05041103}
or Ref.\,\onlinecite{Han0611438},
is now dimensionless and possesses a unified error tolerance
as that of $\rho_0$.
Thus, an efficient on--the--fly filtering algorithm that also
automatically truncates the hierarchy can be applied.\cite{Shi09084105}

 Note that the exact HEOM formalism can be
constructed in principle for arbitrary non--Markovian
dissipation;\cite{Xu07031107,Zhe09164708}
see Appendix for the exact theory of Drude dissipation.
There the bath correlation function $C(t)$ is
expanded in Matsubara exponential series of
the exact fluctuation--dissipation theorem [\Eq{FDT}].
The exact theory is generally expensive,
even with the state--of--the--art numerical filtering algorithm.\cite{Shi09084105}
Apparently, the HQME [\Eq{modZEheom}] is numerically appealing,
much more practical to large systems,
if the effect of involving approximation
can be assessed in advance.
We shall discuss this issue later in \Sec{num} and \Sec{thcomm}.

\subsection{Stochastic Liouville equation description}
\label{com}

 Note that the HQME [\Eq{modZEheom}] has
basically the same mathematical form, and thus
about the same numerical cost, as the
conventional stochastic
Liouville equation.\cite{Tan06082001,Kub63174,Kub69101,Abr092350,Zhu093750}
The latter is based on the classical bath $C_{\rm cl}(t)$
[\Eq{Ctcl}], and can be
recovered from \Eq{modZEheom} by setting $c_i=\Delta=0$.
Thus, the HQME supports the same physical picture,
as described by $\hat\rho(\Omega,t)$ in the stochastic
Liouville equation, with the diffusive
solvation variable $\Omega$ being introduced
for the effect of bath.
The reduced system density operator
is evaluated via $\rho(t)=\int d\Omega\hat\rho(\Omega,t)$.
The HQME [\Eq{modZEheom}]
resolves the stochastic description as
\be\label{eq:rhOme}
  \hat\rho(\Omega, t) = \sum_{n=0}^{\infty} \rho_n(t)\phi_{n}(\Omega)\, ,
\ee
where $\phi_{n}(\Omega)=e^{-\frac{\Omega^2}{4}}\phi^{\rm har}_n(\Omega)$,
with $\phi^{\rm har}_n(\Omega)$ being the normalized harmonic eigenfunction,
is the right--eigenfunction of the
diffusion operator,\cite{Tan89101,Tan06082001}
\be\label{GamOme}
\Gamma_\Omega = -\gamma \frac{\partial}{\partial \Omega}
\left( \Omega + \frac{\partial}{\partial \Omega}\right) .
\ee
With the same algebra of Ref.\,\onlinecite{Shi09164518},
we can show that the HQME [\Eq{modZEheom}] is equivalent to the
following stochastic Liouville equation description,
\begin{align} \label{equiZE}
\frac{\partial}{\partial t}\hat\rho(\Omega,t)
& =- \left(i\mathcal{L}+\delta{\cal R}+\Gamma_{\Omega}\right)\hat\rho(\Omega,t)
\nl&\quad
  - i\sqrt{|c_r|}
  \Big[Q,\big(\Omega+\frac{\partial}{\partial\Omega}\big)\hat\rho(\Omega,t)\Big]
\nl &\quad
  + i\frac{c_r}{\sqrt{|c_r|}}
  \Big[Q,\frac{\partial}{\partial\Omega}\hat\rho(\Omega,t)\Big]
\nl &\quad
  +  \frac{c_i}{\sqrt{|c_r|}}
  \Big\{Q,\frac{\partial}{\partial\Omega}\hat\rho(\Omega,t)\Big\}.
\end{align}
In the classical bath limit (setting $\Delta=c_i=0$), the
above equation reduces to the conventional stochastic Liouville
equation.\cite{Tan06082001,Kub63174,Kub69101,Abr092350,Zhu093750}
We have therefore extended the stochastic Liouville equation
to not just the HTA bath, but also
the present improved semiclassical scheme.
 Moreover, for the spin--boson system,
the HTA ($\Delta=0$) version of \Eq{equiZE}
has been recently shown\cite{Shi09164518} to be identical to
the Zusman equation (ZE).\cite{Zus80295,Zus8329,Gar854491,Yan89281}
Therefore,  \Eq{modZEheom} or \Eq{equiZE}
can also be considered as a generalized and modified ZE to arbitrary systems,
with much improved validity range of parameters;
see \Sec{num} for a thorough demonstration.

\subsection{Continued fraction Green's function formalism}
\label{green}

 The HQME formalism can in general apply
to the systems in the presence of external time--dependent field driving.
In this case, the initial conditions to \Eq{modZEheom} are
the steady--states $\{\rho_n(t=0)=\rho^{\rm st}_{n}\}$
before the external field driving.
This initial conditions can be evaluated by setting $\dot\rho_n=0$,
leading to \Eq{modZEheom} a set of linear equations, under the constraint
of Tr$\rho_0=1$. It results in $\rho_{n>0}(t=0)\neq 0$ generally,
due to the initial system--bath coupling.

 For the population transfer systems
to be studied in the absence of external field driving,
we set the initial state to be
$\rho_n(t=0)=\rho(0)\delta_{n0}$.
This corresponds to the initial total density matrix
factorization ansatz.
In this case, the HQME [\Eq{modZEheom}] can be formally
resolved with a continued fraction Green's function
formalism.\cite{Tan06082001,Tan89101,Tan914131,Han0611438,Xu07031107}
Following the same procedure as Ref.\ \onlinecite{Han0611438},
we introduce the hierarchical Liouville--space
propagators $\{\cU_{n\geq 0}(t)\}$ via
\be \label{Un}
 \rho_n(t)\equiv e^{-n\gamma t}\cU_n(t)\rho(0);
\ \  {\rm with\ \ } \cU_n(0)=\delta_{n0},
\ee
and recast \Eq{modZEheom} as (setting $i\cL'\equiv i\cL+\delta{\cal R}$)
\begin{align}
  \dot\cU_n(t)&=-i\cL' \cU_n(t)
   -i\sqrt{n}\cA e^{\gamma t}\cU_{n-1}(t)
\nl & \quad
  -i\sqrt{n+1}\cB e^{-\gamma t}\cU_{n+1}(t).
\end{align}
In the Laplace domain, we have
\begin{align} \label{Uns}
  \delta_{n0}&=
 (s+i\cL')\ti\cU_n(s)+i\sqrt{n}\cA\ti\cU_{n-1}(s-\gamma)
\nl & \quad
                  + i\sqrt{n+1}\cB\ti\cU_{n+1}(s+\gamma).
\end{align}
Define the hierarchical Liouville--space Green's functions $\{\cG^{(n)}(s)\}$ via
\bsube \label{Gndef}
\begin{align}
  \ti\cU_0(s) &\equiv  \cG^{(0)}(s),
\label{Grn0} \\
  \ti\cU_n(s) &\equiv -i\sqrt{n}\,\cG^{(n)}(s)\cA
  \ti\cU_{n-1}(s-\gamma); \ \ n>0.
\label{Gn}
\end{align}
\esube
Then \Eq{Uns} leads to
\bsube \label{GPi}
\be \label{Gs}
  \cG^{(n)}(s) =\frac{1}{s+i\cL'+\Pi^{(n)}(s)}\, ,
\ee
with
\be \label{Pis}
\Pi^{(n)}(s) \equiv (n+1)\cB\cG^{(n+1)}(s+\gamma)\cA\, .
\ee
\esube
The above equations constitute the continued fraction formalism
to evaluate each individual $\Pi^{(n)}(s)$ or  $\cG^{(n)}(s)$.

 Note that $\cG^{(0)}(s)\equiv\cG(s)$
and its associated $\Pi^{(0)}(s)\equiv\Pi(s)$
are the primary Green's function and dissipation kernel resolution,
respectively.
The reduced density operator $\ti\rho(s)=\cG(s)\rho(0)$ satisfies
\be \label{rhos}
    s\ti \rho(s) - \rho(0) =
   -i\cL'\ti\rho(s)-\Pi(s)\ti\rho(s),
\ee
which in the time domain reads
\be \label{rhot}
  \dot\rho(t) = -i\cL'\rho(t) - \int_0^{t}\!d\tau \,
     \hat\Pi(t-\tau)\rho(\tau).
\ee
This is nonperturbative quantum master equation, with
$\hat\Pi(t-\tau)$ [or $\Pi(s)$] evaluated nonperturbatively via the continued
fraction formalism and $i\cL'\equiv i\cL+\delta{\cal R}$.

 To obtain the kinetic rate equations,
we start with \Eq{rhos} and
formally eliminate the coherence
components of the reduced density matrix, resulting in
\be \label{Pvecs}
       s\ti{\mbox{\boldmath $P$}}(s)
          -{\mbox{\boldmath $P$}}(0)
 = K(s) \ti{\mbox{\boldmath $P$}}(s).
\ee
Its time--domain counterpart reads
\be \label{ratePt0}
 \dot P_j(t)=\sum_k \int_0^t\!d\tau \hat{K}_{jk}(t-\tau) P_k(\tau),
\ee
where $K_{jk}(s)$ are transfer rates resolutions
for the transition from the state $k$ to the state $j$.
They can be obtained via
\be\label{rateKs}
   K(s) =  T_{\mbox{\tiny PC}}(s+T_{\mbox{\tiny CC}})^{-1}
     T_{\mbox{\tiny CP}} - T_{\mbox{\tiny PP}} .
\ee
Here, $T_{\mbox{\tiny PC}}$, $T_{\mbox{\tiny CC}}$,
$T_{\mbox{\tiny CP}}$, and $T_{\mbox{\tiny PP}}$
denote the coherence-to-population, coherence-to-coherence,
population-to-coherence, and population-to-population
transfer matrices, respectively, defined by \Eq{rhos} in
a given representation.

\section{Analytical resolution of electron transfer dynamics}
\label{ana}

 Consider hereafter the standard electron transfer model system.
The total system--plus--bath composite Hamiltonian assumes
$H_{\rm T} = h_a |a\ra\la a| + (h_b+E^{\circ})|b\ra\la b|
 + V(|a\ra\la b|+|b\ra\la a|)$.
Here, $E^{\circ}$ denotes the reaction endothermicity,
and $V$ is the transfer coupling matrix element that is assumed
independent of the solvent degrees of freedom;
$h_a$ or $h_b$ is the
solvent Hamiltonian for the
system in the donor $|a\ra$
or acceptor $|b\ra$ state, respectively.
Their difference defines the solvation coordinate,
\be\label{Udef}
 U \equiv h_b-h_a.
\ee
The system is assumed to be initially in the
donor state, with
$\rho_{\rm T}(0)=|a\ra\la a|\rho_{a}^{\rm eq}$,
where the bath Hamiltonian assumes
$h_{\B}=h_a$; i.e., $\rho_{\B}^{\rm eq}=\rho_{a}^{\rm eq}\propto e^{-\beta h_a}$.
It also defines the solvation reorganization energy
$\lambda =\la U \ra_{\B}={\rm tr}_{\B}(U\rho^{\rm eq}_{\B})$
and the generalized Langevin force
$F_{\B}(t)=e^{ih_{\B}t}(U-\lambda)e^{-ih_{\B} t}$
that physically corresponds to the diffusive
variable $\Omega$ introduced in \Sec{com}.

  The reduced electron transfer system Hamiltonian reads now
\be\label{Hsys}
  H = (E^{\circ}+\lambda)|b\ra\la b| + V(|a\ra\la b|+|b\ra\la a|),
\ee
while the dissipative system mode is
\be\label{Qmode}
 Q = |b\ra\la b|.
\ee
The following derivation of the analytical
solutions to the transfer dynamics follows
the same algebra as Ref.\,\onlinecite{Han0611438},
where the HTA version was treated.
By analyzing the tensor elements involved
in \Eq{Pis} for the specified dissipative mode
$Q$ of \Eq{Qmode}, we find that
the only nonzero tensor elements of $\Pi^{(n)}$ remain
to be
\be \label{xyz}
   x^{(n)} \equiv  \Pi^{(n)}_{ba,ba}, \
   y^{(n)} \equiv  \Pi^{(n)}_{ba,ab}, \
   z^{(n)} \equiv  \Pi^{(n)}_{ba,bb},
\ee
and their Hermitian conjugate counterparts.
They are related to the Green's function tensor elements,
\be \label{XYZ}
   X^{(n)} \equiv  \cG^{(n)}_{ba,ba}, \
   Y^{(n)} \equiv  \cG^{(n)}_{ba,ab}, \
   Z^{(n)} \equiv  \cG^{(n)}_{ba,bb} ,
\ee
via [cf.\ \Eq{Pis} and denoting $\eta\equiv c_r-ic_i$]
\bsube \label{finaldebye}
\bea
   x^{(n)}(s) &=& \eta (n+1) X^{(n+1)}(s+\gamma),
    \ \  \ \
\\
  y^{(n)}(s) &=& -\eta^{\ast}(n+1) Y^{(n+1)}(s+\gamma),
\\
  z^{(n)}(s) &=& (\eta-\eta^{\ast})(n+1) Z^{(n+1)}(s+\gamma).
\eea
\esube
From \Eq{Gs} and the associated Dyson equation method,\cite{Han0611438}
we then obtain
\bsube \label{finalXYZ}
  \begin{align} \label{finalX}
 X^{(n)}(s) &= [\alpha^{(n)}(s)+\beta^{(n)}(s)]^\ast/\zeta^{(n)}(s),
 \\ \label{finalY}
 Y^{(n)}(s) &= [\beta^{(n)}(s)-y^{(n)}(s)]/\zeta^{(n)}(s),
 \\ \label{finalZ}
 Z^{(n)}(s) &= -\frac{1}{s}
   \Bigl\{[z^{(n)}(s)-iV] X^{(n)}(s) \nl & \qquad\
       +  [z^{(n)}(s)-iV]^\ast Y^{(n)}(s)\Bigr\},
\end{align}
\esube
with $\zeta^{(n)}(s)\equiv|\alpha^{(n)}(s)
  + \beta^{(n)}(s)|^2 - |\beta^{(n)}(s) - y^{(n)}(s)|^2$
and
\bsube \label{allaug}
\begin{align}
\alpha^{(n)}(s)  &\equiv  s+i(E^{\circ}+\lambda)+\Delta + x^{(n)}(s),
\label{alpha} \\
 \beta^{(n)}(s)  &\equiv  s^{-1} V [ 2V + iz^{(n)}(s)].
\label{beta}
\end{align}
\esube
The above formulations [\Eqs{xyz}--(\ref{allaug})]
constitute the inverse recursive
analytical evaluation of
the required dissipation kernel
$\Pi(s)\equiv \Pi^{(0)}(s)$, and also the Green's function.

 The forward and backward rate resolutions $k(s)$ and $k'(s)$
for the present two--level system of study can then
be evaluated by using \Eq{rateKs}, resulting in
\bsube\label{finalK}
\be \label{Ks}
  k(s) = 2V^2{\rm Re}\left\{
    \frac{\alpha(s)+y(s)}{|\alpha(s)|^2 - |y(s)|^2}
   \right\},
\ee
and
\be \label{Kbs}
  k'(s) = 2V^2{\rm Re}\left\{
    \frac{[\alpha(s)+y(s)] [1-iz^{\ast}(s)/V]}{|\alpha(s)|^2 - |y(s)|^2}
  \right\}.
\ee
\esube
 The rate constants
are the values at $s=0$ for their steady state nature.
The equilibrium reduced density matrix is obtained
by solving $[i\cL'+\Pi(s=0)]\rho^{\rm eq}=0$, together with Tr$\rho^{\rm eq}=1$.
The solutions with $s=0$ are
\bsube\label{rhoeq}
\begin{align}
\rho^{\rm eq}_{bb}&=1-\rho^{\rm eq}_{aa}=\frac{
     {\rm Re}(\alpha+y)}
      {{\rm Re}[(\alpha+y)(2-iz^{\ast}/V)] },
\\
 \rho^{\rm eq}_{ab}&=\rho^{\rm eq}_{ba}=-\frac{{\rm Re}\, z}
                {{\rm Re}[(\alpha+y)(2-iz^{\ast}/V)]}.
\end{align}
\esube

  Note that the electron transfer system considered here
is different from the standard spin--boson model
by their initial equilibrium bath states.
The former is determined by $h_{\B} = h_a$,
while the latter by $h_{\B}=\frac{1}{2}(h_a+h_b)$.
This distinction results in different
forms of the reduced system Hamiltonian
and dissipative mode; thus different $\rho(t)$
dynamics. The steady--state behaviors
would be the same, as required by thermodynamics
principles,
if the exact QDT is used.
However, as here approximations are involved in treating
the bath correlation function,
the steady state behaviors such as rate constants
and equilibrium system density matrix
can be different in the aforementioned two model
systems.
Nevertheless, for either the standard electron transfer
or spin--boson system, the algebra to
analytical solutions is same.

\section{Numerical validations}
\label{num}

  In this section we shall show
that the proposed HQME is remarkably superior
over the original HTA/ZE scheme.
We address the issues of validity and applicability
by considering the positivity and accuracy of
the reduced density matrix dynamics.
A versatile criterion for the range of applicability
of the HQME
will be constructed later; see the next section.
Numerical results are all reported in
unit of ${k_BT}=1$.

\begin{figure}
\includegraphics[width=0.75\columnwidth]{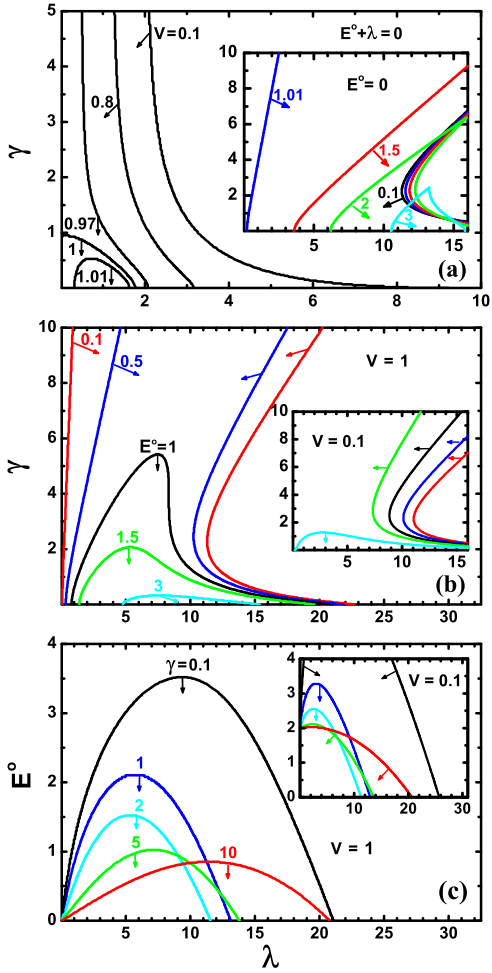}
\caption{Positivity diagram of Zusman equation,
where the P--region is indicated by the arrow or arrows associated with each curve
 or each pair of curves in same color.
(a): Diagram in ($\lambda,\gamma; V$)--subspace (with some selected values of $V$)
for classical barrierless ($E^{\circ}+\lambda=0$)
and symmetric ($E^{\circ}=0$; inset) systems,
(b): Diagram in ($\lambda,\gamma; E^{\circ}$)--subspace for $V=1$ and 0.1 (inset).
(c): Same as (b) but plotted in ($E^{\circ},\lambda;\gamma$)--subspace.
Unit of $k_BT=1$ is used.
}
\label{fig1}
\end{figure}

\begin{figure}
\includegraphics[width=0.85\columnwidth]{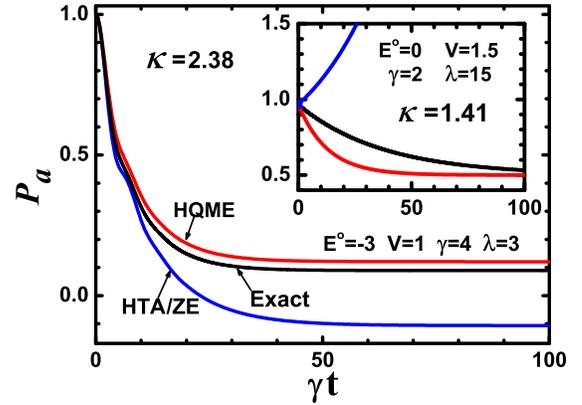}
\caption{Population evolutions: Exact (black),
 HQME (red), and HTA/ZE (blue), for the specified systems.
For the parameter $\kappa$, see \Eq{kappa}.
Unit of $k_BT=1$ is used.
}
\label{fig2}
\end{figure}

\begin{figure}
\includegraphics[width=0.85\columnwidth]{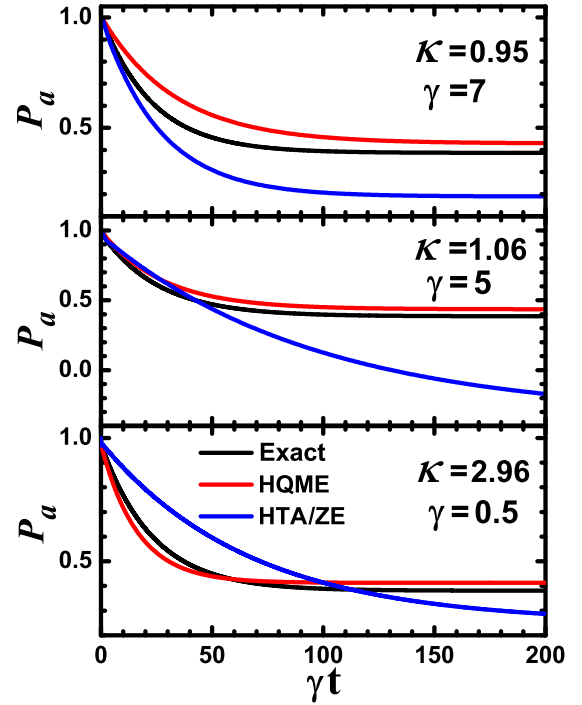}
\caption{Same as \Fig{fig2}, but for systems with
 the common $(E^{\circ},V,\lambda)=(-0.5,1, 13)$,
 and different values of $\gamma$, specified individually
 in each panel.
 Unit of $k_BT=1$ is used.
}
\label{fig3}
\end{figure}

  For the issue of positivity, we
focus on the asymptotic regime, characterized by
the so--called P--region
in the electron transfer parameters space, where
the rate constants, $k\equiv k(s=0)$ and $k\equiv k'(s=0)$,
and the equilibrium density matrix
{\it all} satisfy the positivity requirement.
The latter amounts to
$\rho^{\rm eq}_{aa}\rho^{\rm eq}_{bb}
 \geq\rho^{\rm eq}_{ab}\rho^{\rm eq}_{ba}$.
With the aid of the analytic results presented in
\Eqs{finalK} and (\ref{rhoeq}),
we explore thoroughly the P--regions in the
temperature--scaled parameters
($\lambda,\gamma,E^{\circ},V$)--space
of the electron transfer system.

 The resulting positivity diagrams
of the ZE are reported in \Fig{fig1},
identical for endothermicity
$E^{\circ}$ and $-E^{\circ}$.
The P--region is indicated by the arrow(s)
associated with each curve (or
pair of curves of same color).
In general, the P--region is larger for higher
temperature, as expected.
Also, the P--region itself does not vary
when the transfer coupling strength $\beta V\leq 0.1$,
but changes quite dramatically
as $\beta V$ increases.
Observations for two specific systems, as
depicted in \Fig{fig1}(a) and its inset,
are as follows.
For a classical barrierless ($E^{\circ}+\lambda=0$) system,
the P--region in
($\lambda,\gamma$)--subspace
quickly shrinks with $\beta V>0.1$, and
the ZE violates positivity
completely for $\beta V\geq 1.04$.
 For a symmetric ($E^{\circ}=0$) system,
the P--region does not change much, if $\beta V\leq 1$,
covering over all $\beta\lambda<11$;
however, when $1<\beta V<4.5$, it confines
only within certain range of $\beta\lambda$
depending on the value of $\beta\gamma$.
The ZE violates positivity completely for $V>4.5$
for a symmetric system.
Apparently, the applicability range of
HTA/ZE scheme depends
not just  on bath interaction parameters,
but also sensitively on system itself,
leading to the complicated P--region diagrams,
as depicted in \Fig{fig1}.

 The HQME  is simply superb.
While the ZE is subject to severe positivity violation especially
for $\beta V>1$,
the new scheme is found to have
P--region covering over
a broad range of parameters space, as tested:
$\beta\lambda<25$, $\beta\gamma<100$, $|\beta E^{\circ}|<25$,
and $\beta V<100$.
In other words, the HQME scheme preserves positivity, at least asymptotically,
in almost entire parameters
space of practical interest.
Moreover, it is likely to support
a convenient and system--insensitive
criterion for its range of applicability; see
the next section.

 The HQME shows also its superiority in time evolution.
It follows the exact results
closer than the HTA/ZE does.
Figure \ref{fig2} depicts
the transfer population evolutions
of HQME, ZE, and exact HEOM.
Chosen here are a classical barrierless
system ($E^{\circ}+\lambda=0$)
in the main panel and a symmetric system
($E^{\circ}=0$) in the inset.
In contrast to the ZE that
evolves into unphysical regime of positivity violation,
the HQME remains accurate quantitatively
and semi--quantitatively, respectively,
for the two specified systems in study.
Figure \ref{fig3} is for
three other systems,
with common $\beta V$, $\beta\lambda$,
and $\beta E^{\circ}$,
but different values of $\beta\gamma$,
where the intermediate one ($\beta\gamma=5$)
leads to the ZE dynamics violation of positivity.
 In all cases studied, the HQME performs
much better than the HTA/ZE does.
The superiority can be qualitative.
While the HTA/ZE violates the positivity,
the HQME can remain even quantitatively applicable.

\section{Discussion and comments}
\label{thcomm}

\subsection{Criterion and measure on applicability}
\label{thcommA}

\begin{figure}
\includegraphics[width=0.85\columnwidth]{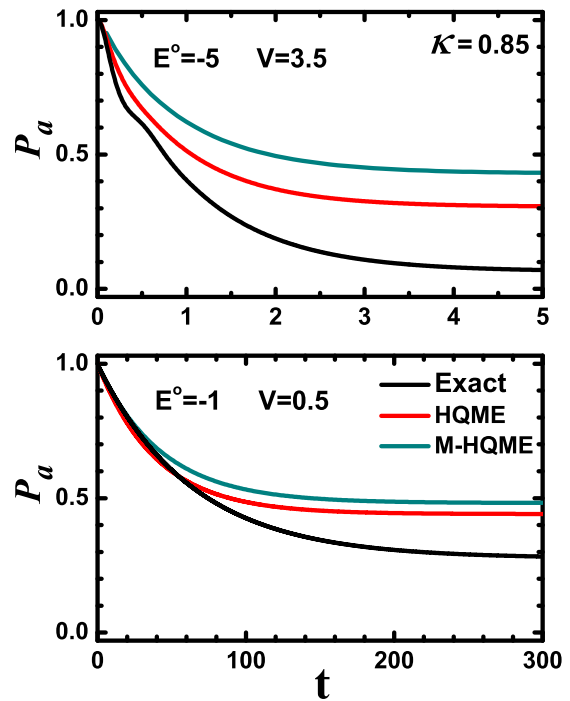}
\caption{Population evolutions: Exact (black),
 HQME (red), and M-HQME (green), for
 $(E^{\circ},V) = (-1,0.5)$ (lower) and $(E^{\circ},V) =(-5,3.5)$ (upper),
 respectively, with same bath parameters of $(\lambda,\gamma)=(20,5)$
 that result in $\kappa=0.85$.
 Both systems disqualify the HTA/ZE for its violating positivity.
 Unit of $k_BT=1$ is used.
}
\label{fig4}
\end{figure}

\begin{figure}
\includegraphics[width=0.85\columnwidth]{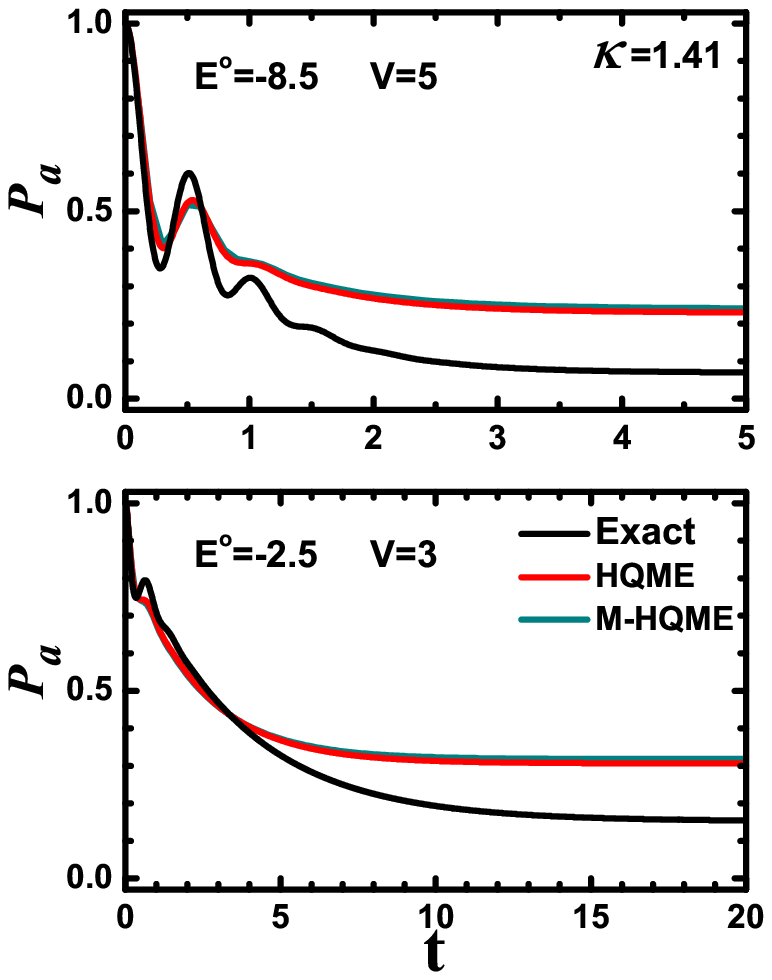}
\caption{Same as \Fig{fig4}, but for
 $(E^{\circ},V) = (-2.5,3)$ (lower) and $(E^{\circ},V) =(-8.5,5)$ (upper),
 respectively, with same bath parameters of $(\lambda,\gamma)=(15,2)$
 that result in $\kappa=1.41$ (same as the inset of \Fig{fig2}).
 Both systems disqualify the HTA/ZE for its violating positivity.
 Unit of $k_BT=1$ is used.
}
\label{fig5}
\end{figure}

\begin{figure}
\includegraphics[width=0.85\columnwidth]{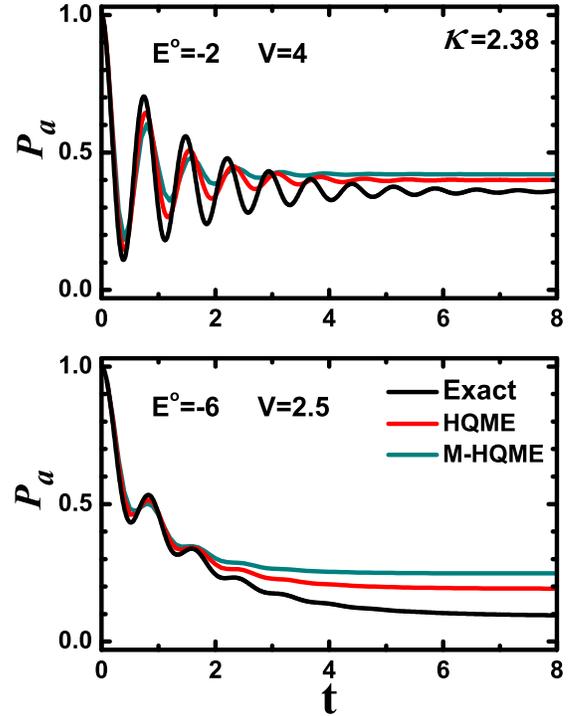}
\caption{Same as \Fig{fig4}, but for
 $(E^{\circ},V) = (-6,2.5)$ (lower) and $(E^{\circ},V) =(-2,4)$ (upper),
 respectively, with same bath parameters of $(\lambda,\gamma)=(3,4)$
 that result in $\kappa=2.38$ (same as the main panel of \Fig{fig2}).
 Both systems disqualify the HTA/ZE for its violating positivity.
 Unit of $k_BT=1$ is used.
}
\label{fig6}
\end{figure}

 For the issue of applicability,
we propose to use
\bsube\label{kappa}
\be\label{kappa0}
 \kappa = \sqrt{6\Gamma(\gamma)/(\beta\lambda\gamma)} ,
\ee
with
\be\label{hwhmapp}
 \beta\Gamma(\gamma)= 6+\sqrt{12+(\beta\gamma)^2} \  ,
\ee
\esube
to estimate both the range of applicability
(setting to be $\kappa>1$)
and the quality of HQME dynamics.
The HQME dynamics tends to be more accurate
for a larger $\kappa$ case.
Justifications on the above criterion
will be made in the next subsection.

 We had indicated in \Fig{fig2} and \Fig{fig3}
their associated values of the
$\kappa$ parameter.
The above system--independent criterion
based on $\kappa$
is shown to be overall satisfactory.
A quantitative agreement of HQME with the exact HEOM dynamics
would depend on the specific details of systems.
However, the system dependence is relatively insensitive,
comparing to that on the nature of bath characterized
by the parameter $\kappa$.

 To further demonstrate this property
of the HQME, we report in \Figs{fig4}--\ref{fig6}
the results of transfer dynamics for
three pairs of systems that are chosen rather arbitrarily.
Two systems in a pair share
a common ($\lambda,\gamma$),
and thus, are associated with
one value of $\kappa$, as indicated in each figure.
These results all indicate that
the proposed HQME supports
a broad range of applicability,
which can be fairly well described by the
quality parameter $\kappa$ of \Eq{kappa}.

 All six cases presented in \Figs{fig4}--\ref{fig6}
are chosen to have the HTA/ZE unphysical, falling
outside of its P--region, cf.\ \Fig{fig1}.
The corresponding HTA/ZE dynamics
are therefore all wrong qualitatively
and not shown.
Rather we compare the results with
an alternative HQME construction,
labeled as M-HQME. This alternative
is based on a similar approximation of
the bath correlation function, but treated
on the Matsubara series expansion of \Eq{eqCt}.
The resulting M-HQME differs from
the present HQME only by the values
of parameters $c_r$ and $\Delta$.
Instead of \Eq{cDel}, the M-HQME assumes
$c_r={\rm Re}\,c_0=\lambda\gamma\cot(\frac{\beta\gamma}{2})$
and $\Delta=\Delta_0=\lambda[\frac{2}{\beta\gamma}
 -\cot(\frac{\beta\gamma}{2})]$, from \Eqs{eqc0}
and (\ref{DeltaK}), respectively.
Apparently, the present HQME [\Eq{modZEheom} with \Eq{cDel}], not just removes the
singularity of cotangent function,
but also performs better in dynamics, as compared to the exact results.

\subsection{Justification of the quality parameter}
\label{thcommB}

\begin{figure}
\includegraphics[width=0.85\columnwidth]{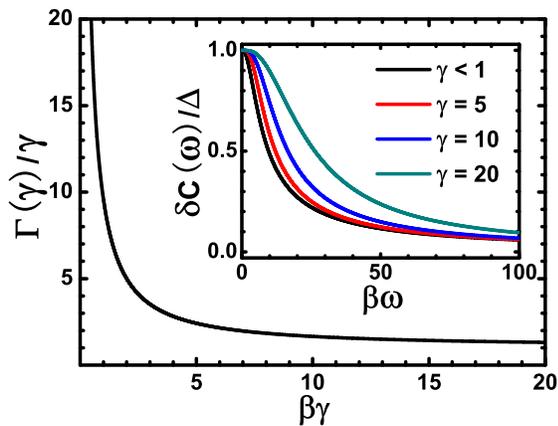}
\caption{The scaled half-width-half-maximum, ${\Gamma(\gamma)}/{\gamma}$,
as function of $\beta\gamma$. The approximate curve via
\Eq{hwhmapp} is found indistinguishable from the accurate
one via \Eq{hwhm}.
Shown in the inset are
$\delta C(\omega)/\Delta$ [cf.\ \Eq{delCw}] versus $\beta\omega$,
at some selected values of $\gamma$ (unit of $k_BT$).
The curve does not change for $\beta\gamma<1$.
}
\label{fig7}
\end{figure}

\begin{figure}
\includegraphics[width=0.85\columnwidth]{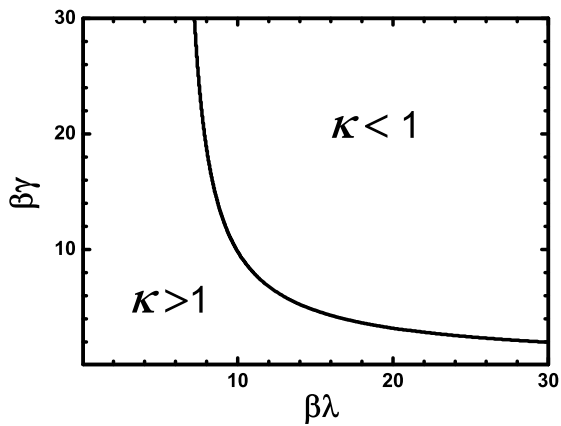}
\caption{Validity diagram of HQME by the criterion of
 the parameter $\kappa$ [\Eq{kappa}].}
\label{fig8}
\end{figure}

 Let us start with some insights
for the qualitative superiority of HQME over HTA/ZE.
Note that the only approximation involved in
each of them individually
is the semiclassical bath correlation function $C_{\rm sc}(t)$ [\Eq{CtMZE}]
versus its high--temperature version, $C_{\HTA}(t)$ [\Eq{CtHTA}],
respectively.
However, the resulting HQME acquires
qualitatively distinct modifications,
especially with $i{\cal L}'\equiv i{\cal L}+\delta{\cal R}$
appearing in \Eq{modZEheom},
that lead to its being at least qualitatively consistent with the exact theory [\Eq{heom}].
Physically, $\delta{\cal R}$ serves as
a diffusion correction to the effective
system Liouvillian, due to the approximated
treatment of bath correlation function.
The HTA/ZE scheme, where $\delta{\cal R}=0$,
does not have this diffusion modification to the system Liouvillian.
Consequently, the applicability range of
HTA/ZE depends
not just  on bath
correlation function $C_{\HTA}(t)$,
but also sensitively on system.

 The HQME dynamics that includes explicitly
the diffusion modified ${\cal L}'$ may thus
support a system--insensitive
criterion for its range of applicability.
This argument has in fact been verified
extensively in our numerical study of transfer systems.
The applicability of HQME may therefore be
addressed by examining the approximation involved
in $C_{\rm sc}(t)$.
We quantify the approximation with
the discrepancy function in the frequency--domain,
\begin{align}\label{delCw}
  \delta C(\omega)
&= C(\omega)-[C_{\rm sc}(\omega)-\Delta]
\nl&= J(\omega)\left(
  \frac{1}{1-e^{-\beta\omega}}-\frac{1}{\beta\omega}
   - \frac{1}{2}-\frac{\beta\omega}{12}
    \right)+\Delta
\nl&= \frac{\lambda\gamma\omega}{\omega^2 + \gamma^2} \Bigl(
     \coth\frac{\beta\omega}{2}-\frac{2}{\beta\omega}
     +\frac{\beta\gamma^2}{6\omega}
     \Bigr).
\end{align}
The Drude model of \Eq{DrudeJ} and
$\Delta=\beta\lambda\gamma/6$ are used explicitly
in the last identity.
The discrepancy function is positive, symmetric, and monotonic
decreasing, with the limiting values of
$\delta C(\omega =0)=\Delta$ and $\delta C(\omega\rightarrow\infty)=0$.
It is demonstrated in the inset of \Fig{fig7}.
Now let $\Gamma(\gamma)$ be the half--width at half--maximum,
determined via
\be\label{hwhm}
  \delta C(\omega)\vert_{\omega=\Gamma(\gamma)}= \frac{\Delta}{2}
  =\frac{\beta\lambda\gamma}{12} .
\ee
It can be well approximated by \Eq{hwhmapp}.
The resulting $\Gamma(\gamma)/\gamma$,
as participated in \Eq{kappa}, is
demonstrated in \Fig{fig7},
with no visible difference from the numerically exact
evaluation of \Eq{hwhm}.

 The criterion on applicability of HQME
can now be considered for the condition under which
$\delta C(\omega)$ can be treated as
Markovian white noise for its effect
of $\delta{\cal R}$ on dissipative systems.
We adopt the Kubo's modulation parameter\cite{Kub63174,Kub69101}
for the criterion:
 $\kappa =   \sqrt{\Gamma(\gamma)/\Delta} = \sqrt{6\Gamma(\gamma)/(\beta\lambda\gamma)} > 1$.
This is the quality parameter of \Eq{kappa}.

 The range of applicability of HQME, as depicted by
the region of $\kappa>1$ in \Fig{fig8},
is quite impressive.
Consider, for example, the fact that
it covers the value of $\beta\lambda<57$, when
$\beta\gamma<1$.
For a rather fast solvation of $\gamma^{-1}=100$ fs,
that $\beta\gamma<1$ would support temperature $T> 75$ K
and reorganization energy at least the range of $\lambda<3000$ cm$^{-1}$.
This covers almost all systems of practical interest.
Note that all dynamics results presented
in this work are in the strong system--bath coupling regime.
Let $\alpha$ be the dimensionless system--bath coupling strength parameter,
defined via $[J(\omega)/\omega]_{\omega=0}\equiv\pi\alpha/2$,
as the Kondo parameter in the Ohmic model.
We have $\alpha=4\lambda/(\pi\gamma)$.
The present HQME would support $\alpha < 72$,
with $\gamma^{-1}=100$ fs and $T=75$ K,
if $\kappa>1$ criterion is used.
The calculations
demonstrated in this paper have the coupling strength
ranging from $\alpha=0.95$ (\Fig{fig2} and
\Fig{fig6})
to $\alpha=33$ (the lower panel of \Fig{fig3}).

\section{Concluding remarks}
\label{sum}

  We have proposed the HQME that may be used
with good confidence to arbitrary systems.
The theoretical development is rooted
in an improved semiclassical
treatment of Drude bath.
This alone improves the conventional
high--temperature or classical approximation
by {\it two or three orders} in temperature parameter,
respectively,
as argued in \Sec{thintro}.
The resulting HQME can be considered
as a natural extension and modification of the conventional
stochastic Liouville equation and Zusman equation.
While it retains their appealing physical pictures
and numerical efficiency,
the HQME shows remarkable improvement over
the conventional theories.
 Its broad range of validity and applicability,
in terms of density matrix positivity and
dynamics quality, are extensively
demonstrated on two--level model systems.
We have also proposed a criterion to estimate the performance
of HQME.
This criterion depends only on the system--bath coupling strength,
characteristic bath memory time and temperature.
Our results all reveal that the HQME
may serve as a versatile tool wherever
the exact approach is
numerically too expensive to afford.

\acknowledgments
 Support from the NNSF China (20533060,
20773114, 20733006 and 20873157),
 the National Basic Research Program of China (2006CB922004),
the CAS China
(KJCX2.YW.H17 and the Hundred Talents Project), and the RGC Hong Kong
SAR Government (604508 and 604709) is acknowledged.

\appendix*
\section{Exact HEOM formalism}
The HEOM formalism for the
reduced system dynamics can be constructed using the
calculus--on--path--integral
technique.\cite{Tan89101,Tan906676,Tan914131,Tan9466,%
Ish053131,Xu05041103,Xu07031107,%
Jin08234703,Zhe09164708}
 This formalism is in principle exact and
nonperturbative. However, its specified form depends on the way of
treating the bath correlation function, under the constraint
of the exact fluctuation--dissipation theorem [\Eq{FDT}].
 Consider the Matsubara expansion method,
in which the bath correlation function
with the Drude model of \Eq{DrudeJ} reads
\be\label{eqCt}
  C(t>0) = c_0 e^{-\gamma t} + \sum_{k=1}^\infty c_ke^{-\gamma_k t} .
\ee
The first term, with
\be\label{eqc0}
  c_0 = \lambda \gamma\left[\cot\bigl(\beta\gamma/{2}\bigr)-i\right] ,
\ee
arises from the Drude pole. The sum term arises
from the Matsubara poles or frequencies of $\gamma_{k\geq 1}=2k\pi/\beta$,
with
\be\label{eqck}
  c_k=\frac{8k\pi\lambda\gamma}{(2k\pi)^2-(\beta\gamma)^2} \,  ;  \  \   k \geq 1   \, .
\ee

 To construct the HEOM, the infinite sum over the $k$ index
in \Eq{eqCt} need to be truncated. To that end, let $\gamma_0 \equiv \gamma$
and recast \Eq{eqCt} by
\be \label{Ctfm}
  C(t)=\sum_{k=0}^{K} c_k e^{-\gamma_k t}+ 2\Delta_K \delta(t),
\ee
where
\be\label{DeltaK}
 \Delta_K \equiv \sum_{k=K+1}^{\infty}
\frac{c_{k}}{\gamma_{k}}
    = \lambda \left[\frac{2}{\beta\gamma}-\cot\bigl(\frac{\beta\gamma}{2}\bigr)\right]
         -  \sum_{k=1}^{K} \frac{c_{k}}{\gamma_{k}} \, .
\ee
This treatment is in principle exact if the $K$ is chosen large
enough and the resulting reduced
system density matrix dynamics of primary interest is converged.

The resulting HEOM formalism
reads\cite{Ish053131,Xu07031107,Shi09164518,Shi09084105}
\begin{align}\label{heom}
\dot\rho_{\bm n} & = - \Bigl( i{\cal L} +\delta{\cal R}_K +
      \sum_{k=0}^K n_k \gamma_k \Bigr) \rho_{\bm n}
\nl &\quad
    - {i}\sum_{k=0}^K  \sqrt{(n_k+1)|c_k|}\, \bigl[Q,\rho_{{\bm n}_k^+}\bigr]
\nl &\quad
    - {i} \sum_{k=0}^K  \sqrt{n_k/|c_k|} \Bigl(
    c_k Q \rho_{{\bm n}_k^-} - c_k^\ast \rho_{{\bm n}_k^-} Q  \Bigr) \, ,
\end{align}
with
\be\label{delRk}
  \delta\mathcal{R}_K\hat O = \Delta_K [Q,[Q,\hat O]].
\ee
The reduced density operator of primary interest
is $\rho\equiv \rho_{\bm 0}$.
The subscript ${\bm n}=\{n_k\!\geq\!0;\ k=0,
\cdots\!,K\}$, which consists of a set of nonnegative indices,
specifies in general a given auxiliary density operator (ADO)
of $\rho_{\bm n}\equiv \rho_{n_0\cdots n_K}$.
The subscript ${\bm n}_k^\pm$ differs from ${\bm n}$ only by
changing the specified $n_k$ to  $n_k \pm 1$. An $N^{\rm th}$--tier
ADO is of $n_0+\cdots + n_K = N$, and the total number of ADOs at
this tier is $\frac{(N+K)!}{N!\,K!}$. Each $\rho_{\bm n}$ in
\Eq{heom} has been scaled individually to have a uniform dimension
and error tolerance. This validates a simple, on--the--fly,
filtering algorithm for efficient numerical propagation, including
automatic truncation of hierarchy.\cite{Shi09164518,Shi09084105}
However, the exact HEOM calculations is still numerically expensive,
especially for complex systems, under strong coupling with bath at
low temperature.
\clearpage
\newpage

\end{document}